\documentclass[conference]{IEEEtran}
\IEEEoverridecommandlockouts
\usepackage{cite}
\usepackage{amsmath,graphicx}
\usepackage{amssymb,amsfonts}
\usepackage{graphicx}
\usepackage{textcomp}
\usepackage{xcolor}
\usepackage{subfigure}
\usepackage{textcomp} 
\usepackage{siunitx}  
\usepackage{algorithm}
\usepackage{algpseudocode}
\usepackage{xpatch}
\usepackage{pgfplots}
\usepackage{pgfplotstable}
\usepackage{readarray}
\usepackage{siunitx}
\usepackage{bm}
\usepackage[nolist]{acronym}
\usepackage{bbm}
\usepackage{url}
\definecolor{TUMBeamerYellow}    {rgb} {1.000,0.706,0.000}    
\definecolor{TUMBeamerOrange}    {rgb} {1.000,0.502,0.000}    
\definecolor{TUMBeamerRed}       {rgb} {0.898,0.204,0.094}    
\definecolor{TUMBeamerDarkRed}   {rgb} {0.792,0.129,0.247}    
\definecolor{TUMBeamerBlue}      {rgb} {0.000,0.600,1.000}    
\definecolor{TUMBeamerLightBlue} {rgb} {0.255,0.745,1.000}    
\definecolor{TUMBeamerGreen}     {rgb} {0.569,0.675,0.420}    
\definecolor{TUMBeamerLightGreen}{rgb} {0.710,0.792,0.510}    

\DeclareMathOperator{\tr}{tr}
\DeclareMathOperator{\T}{T}
\DeclareMathOperator{\He}{H}

\DeclareMathOperator{\inv}{-1}

\DeclareMathOperator{\diag}{diag}
\DeclareMathOperator{\rank}{rank}

\newcommand{\eye}{\bm{\mathrm{I}}}
\newcommand{\bbthet}{\bm{\bar{\theta}}}
\newcommand{\bD}{\bm{D}}
\newcommand{\Her}{{\He}}
\newcommand{\td}{{\text{d}}}

\newcommand{\tdk}{{\text{d},k}}
\newcommand{\bT}{\bm{T}}
\newcommand{\bC}{\bm{C}}
\newcommand{\bQ}{\bm{Q}}
\newcommand{\bW}{\bm{W}}

\newcommand{\NB}{N_{\text{B}}}
\newcommand{\NM}{N_{\text{M}}}
\newcommand{\NRe}{N_{\text{R}}}
\newcommand{\bPhi}{\bm{\Phi}}
\newcommand{\bLam}{\bm{\Lambda}}

\newcommand{\tre}{{\text{r}}}

\newcommand{\trek}{{\text{r},k}}
\newcommand{\ts}{{\text{s}}}
\newcommand{\bH}{\bm{H}}

\newcommand{\mThet}{\bm{\Theta}}
\newcommand{\bu}{\bm{u}}

\newcommand{\bv}{\bm{v}}
\newcommand{\bthet}{\bm{\theta}}

\newcommand{\cmplx}[1]{\mathbb{C}^{#1}}
\newcommand{\norm}[1]{\|#1\|}
\newcommand{\abs}[1]{|#1|}

\newcommand{\gaussdist}[2]{\mathcal{N}_{\mathbb{C}}(#1,#2)}
\newcommand{\summe}[2]{\sum_{#1}^{#2}}
\newcommand{\produkt}[2]{\prod_{#1}^{#2}}

\def\BibTeX{{\rm B\kern-.05em{\sc i\kern-.025em b}\kern-.08em
    T\kern-.1667em\lower.7ex\hbox{E}\kern-.125emX}}
\begin{document}

\begin{acronym}
    \acro{AoD}{angle of departure}
    \acro{AoA}{angle of arrival}
    \acro{ULA}{uniform linear array}
    \acro{CSI}{channel state information}
    \acro{LOS}{line of sight}
    \acro{EVD}{eigenvalue decomposition}
    \acro{BS}{base station}
    \acro{MS}{mobile station}
    \acro{mmWave}{millimeter wave}
    \acro{DPC}{dirty paper coding}
    \acro{RIS}{reconfigurable intelligent surface}
    \acro{AWGN}{additive white gaussian noise}
    \acro{MIMO}{multiple-input multiple-output}
    \acro{UL}{uplink}
    \acro{DL}{downlink}
    \acro{OFDM}{orthogonal frequency-division multiplexing}
    \acro{TDD}{time-division duplex}
    \acro{LS}{least squares}
    \acro{MMSE}{minimum mean square error}
    \acro{SINR}{signal to interference plus noise ratio}
    \acro{OBP}{optimal bilinear precoder}
    \acro{LMMSE}{linear minimum mean square error}
    \acro{MRT}{maximum ratio transmitting}
    \acro{M-OBP}{multi-cell optimal bilinear precoder}
    \acro{S-OBP}{single-cell optimal bilinear precoder}
    \acro{SNR}{signal-to-noise ratio}
    \acro{SDR}{Semidefinite Relaxation}
    \acro{SE}{spectral efficiency}
    \acro{GCEs}{Gram channel eigenvalues}
\end{acronym}

\title{High SNR Analysis of RIS-Aided\\ MIMO Broadcast Channels\\}

\author{\IEEEauthorblockN{Dominik Semmler, Michael Joham, and Wolfgang Utschick}
\IEEEauthorblockA{\textit{School of Computation, Information and Technology, Technical University of Munich, 80333 Munich, Germany} \\
email: \{dominik.semmler,joham,utschick\}@tum.de}
}

\maketitle

\begin{abstract}
    We analyze the influence of a \ac{RIS} on the Gram channel eigenvalues in a high \ac{SNR} scenario.
    This allows to connect specific channel properties with the rank improvement capabilities of the RIS.
    In particular, fundamental limits due to a possible \ac{LOS} setup between the \ac{BS} and the \ac{RIS} are derived.
    Furthermore, \ac{DPC} based schemes are compared to linear precoding in such a scenario and it is shown that under certain channel conditions, the performance gap between \ac{DPC} and linear precoding vanishes.
\end{abstract}


\section{Introduction}
\begin{figure}[b]
    \onecolumn
    \centering
    \scriptsize{This work has been submitted to the IEEE for possible publication. Copyright may be transferred without notice, after which this version may no longer be accessible.}
    \vspace{-1.3cm}
    \twocolumn
\end{figure}
\label{sec:intro}
RISs have drawn a lot of attention recently as they are viewed as a key technology for future communication systems (e.g., \cite{Power_Min_IRS}). An \ac{RIS} is a surface consisting of many passive low-cost elements which make it possible to enhance the propagation environment.
Under perfect \ac{CSI}, which is also assumed in this article, already large improvements in the power consumption (see \cite{Power_Min_IRS}), the energy efficiency (see \cite{EnergyEff}), and the \ac{SE} (see \cite{WSR}) could be obtained when considering an \ac{RIS} in the communication system.
Specifically for \ac{MIMO} systems, which are considered in this article, the algorithms of \cite{MIMOP2PCap}, \cite{WMMSEMIMO},  \cite{MaxSumRateJour}, \cite{IRSLISA} are available which show a clear improvement when incorporating an \ac{RIS}.\\
Even though the performance can be improved considerably, the phase optimization is typically not intuitive due to the highly non-convex structure of the underlying problem.
Additionally, it is still not entirely clear under which specific channel conditions we can expect a large improvement of the data transmission.
In \cite{RankImprov}, the rank-improvement capabilities of an \ac{RIS} were demonstrated and give the impression that especially for these rank-improvement scenarios a large performance gain can be expected.\\
Changing the effective rank of the channel matrix is directly connected with a change of its corresponding singular values.
We analyze the behavior of the squared singular values which are additionally the eigenvalues of the Gram channel matrix and are referred to as \ac{GCEs} throughout this article.
For conventional communication systems, the modification of the \ac{GCEs} was impossible as the channel could not be manipulated.
When considering \ac{RIS}s, however, an additional channel component via the \ac{RIS} adds up to the direct channel.
The \ac{GCEs} of the resulting composite channel are, therefore, different from the ones of the direct channel and can be modified by the phase optimization of the \ac{RIS}.
This modification of the composite channel \ac{GCEs} in comparison to the direct channel \ac{GCEs} is analyzed in this article.
Moreover, we discuss the limitations and the resulting consequences on the performance of the \ac{MIMO} system. Considering the high \ac{SNR} region allows to directly transfer the observations to the \ac{SE}. 
One particular limitation results from the fact that the \ac{RIS} and the \ac{BS} are typically positioned in \ac{LOS} resulting in the \ac{BS}-\ac{RIS} channel having a low rank.
If no direct channel were present, the composite channel would only consist of the channel via the \ac{RIS} and the low-rank structure directly poses limitations on the composite channel.
However, we will see that, despite a direct channel being present, the eigenvalue modification remains to be severly limited in this particular case. 
In summary, we make the following contributions in this article:
\begin{itemize}
    \item Analysis of the Gram channel eigenvalue modification performed by the \ac{RIS} and its implications on the \ac{SE} at high \ac{SNR}.
    \item Derivation of limitations evolving from the often used \ac{LOS} assumption of the channel between the \ac{BS} and the \ac{RIS}.
    A multi-\ac{RIS} scenario is motivated as a possible solution to these limitations.
    \item We show that, under certain conditions on the channel between the \ac{BS} and the \ac{RIS}, all Gram channel eigenvalues become similar resulting in orthogonal user channels.
    Therefore, the gap between \ac{DPC} based methods (see \cite{MaxSumRateJour}) and the earlier proposed linear scheme (see \cite{IRSLISA}) vanishes for an increasing number of \ac{RIS} elements.
\end{itemize}

\section{System Model}
We consider a scenario with an $\NB$ antenna \ac{BS} serving $K$ users each having $\NM$ antennas.
Moreover, an \ac{RIS} with $\NRe$ reflecting elements is available to enhance the transmission.
The downlink channel from the \ac{BS} to the $k$-th user is therefore given by 
\begin{equation}
    \bH_k = \bH_\tdk + \bH_\trek \mThet \bH_\ts \; \in \cmplx{\NM \times \NB}
\end{equation}
where $\bH_\tdk \in \cmplx{\NM \times \NB}$ is the direct channel from the \ac{BS} to the $k$-th user,  $ \bH_\trek \in \cmplx{\NM \times \NRe}$ is the reflecting channel from the \ac{RIS}
to user $k$, $\mThet = \diag(\bthet) \in \cmplx{\NRe}$ with $\bthet \in \{\bm{z} \in \cmplx{\NRe}: \abs{z_n}=1,\; \forall n\}$ is the phase manipulation at the \ac{RIS}
and $ \bH_\ts \in \cmplx{\NRe \times \NB}$ is the channel from the \ac{BS} to the \ac{RIS}.
The stacked channel matrix, containing the channels for all the users, is defined as 
\begin{equation}
    \bm{H} = \begin{bmatrix}
                    \bm{H}^\Her_1,\bm{H}^\Her_2,\hdots,\bm{H}^\Her_K\\
                \end{bmatrix}^\Her 
                \in \cmplx{ r \times \NB}
\end{equation}
with $r=K\NM $ where we additionally assume $\NB \ge r$.
Also the subchannels $ \bH_\tdk$ and $\bH_\trek $ are stacked accordingly resulting in $\bH_\td \in \cmplx{ r \times \NB}$ and $\bH_\tre\in \cmplx{ r \times \NRe}$.
\section{High SNR Expressions}
As performance metric, we consider the \ac{SE} which for the high SNR discussion can be written (see e.g. \cite{HighSNRInstant}) as
\begin{equation}
    \begin{aligned}
        R_{\text{DPC/Lin}}  &= r\log_2P - r\log_2r + R_{\text{DPC/Lin,Offset}}\\
    \end{aligned}
\end{equation}
with
\begin{equation}\label{eq:HSNROffsets}
    \begin{aligned}
        R_{\text{DPC,Offset}}  &= -\log_2\det{\left((\bm{H}\bm{H}^{\He})^{\inv}\right)},\\
        R_{\text{Lin,Offset}}  &= -\summe{k=1}{K}\log_2\det\left({\bm{E}_k^{\T}(\bm{H}\bm{H}^{\He})^{\inv}\bm{E}_k}\right)
    \end{aligned}
\end{equation}
for \ac{DPC} and linear precoding, respectively.
Therefore, the slope w.r.t. the logarithmic power is the same for both schemes and they are only differing in their offsets.
In \cite{HighSNRInstant}, it has already been discussed that 
$R_{\text{DPC,Offset}} \ge R_{\text{Lin,Offset}}$,
with equality for a blockdiagonal $\bH \bH^\Her$ (orthogonal user channels).
Defining the \ac{EVD} $\bH \bH^\Her = \bT \bLam \bT^\Her$, 
the offset of DPC $R_{\text{DPC,Offset}}$ can be equivalently stated as
$R_{\text{DPC,Offset}} = r\log_2  \overline{\lambda_{\text{G}}}$
where $\overline{\lambda_{\text{G}}}$ is the geometric mean of the \ac{GCEs}.
By applying two times the inequality between the geometric and the arithmetic mean in $(a)$ and $(b)$, it is possible to obtain
\begin{equation}\label{eq:HarmonicMeanBound}
    \begin{aligned} 
    -R_{\text{Lin, Offset}}       &= \summe{k=1}{K}\log_2\det\left(\bm{E}_k^{\T}(\bm{H}\bm{H}^{\He})^{\inv}\bm{E}_k\right)\\
            &= \summe{k=1}{K}N_{\text{M}}\log_2{\det\left(\bm{E}_k^{\T}(\bm{H}\bm{H}^{\He})^{\inv}\bm{E}_k\right)}^{\frac{1}{N_{\text{M}}}}\\
            &\overset{(a)}{\le}  \summe{k=1}{K} N_{\text{M}}\log_2\frac{1}{N_{\text{M}}} {\tr\left(\bm{E}_k^{\T}(\bm{H}\bm{H}^{\He})^{\inv}\bm{E}_k \right)}\\
            &=   N_{\text{M}}\log_2 \produkt{k=1}{K} \frac{1}{N_{\text{M}}}{\tr\left(\bm{E}_k^{\T}(\bm{H}\bm{H}^{\He})^{\inv}\bm{E}_k \right)}\\
            &=   N_{\text{M}}K\log_2 \sqrt[K]{ \produkt{k=1}{K}\frac{1}{N_{\text{M}}} {\tr\left(\bm{E}_k^{\T}(\bm{H}\bm{H}^{\He})^{\inv}\bm{E}_k \right)}}\\
            &\overset{(b)}{\le}   N_{\text{M}}K\log_2 \frac{1}{K} \summe{k=1}{K} \frac{1}{N_{\text{M}}} {\tr\left(\bm{E}_k^{\T}(\bm{H}\bm{H}^{\He})^{\inv}\bm{E}_k \right)}\\
            &=   r\log_2 \frac{1}{r} {\tr\left((\bm{H}\bm{H}^{\He})^{\inv} \right)}\\
            &= -  r\log_2 \frac{r}{\tr\left((\bm{H}\bm{H}^{\He})^{\inv}\right)}\\
            &= -  r\log_2 \overline{\lambda_{\text{H}}}
        \end{aligned}
\end{equation}
and the offset $ R_{\text{Lin, Offset}}$ can, therefore, be bounded as
$r\log_2 \overline{\lambda_{\text{H}}} \le R_{\text{Lin, Offset}} \le r\log_2  \overline{\lambda_{\text{G}}} \le  R_{\text{DPC, Offset}}$
where $ \overline{\lambda_{\text{H}}} $
is the harmonic mean of the eigenvalues.\\
Using this interpretation, we can bound the difference of \ac{DPC} and linear precoding in the high SNR region with the \ac{GCEs} as 
$ R_{\text{DPC}} -  R_{\text{Lin}} \le r\log_2 \frac{\overline{\lambda_{\text{G}}}}{\overline{\lambda_{\text{H}}}}$ which follows from \eqref{eq:HarmonicMeanBound}.
With the help of the \ac{RIS}, we are now able to manipulate the channel and its corresponding \ac{GCEs}.
From \eqref{eq:HarmonicMeanBound}, it follows that the difference of \ac{DPC} and linear precoding can be expressed as
\begin{equation}
    \begin{aligned}
         R_{\text{DPC}}(\bm{\theta}_{\text{Geo}}) -  R_{\text{Lin}}(\bm{\theta}_{\text{Har}}) \le  r\log_2 \frac{\overline{\lambda_{\text{G}}}(\bm{\theta}_{\text{Geo}})}{\overline{\lambda_{\text{H}}}(\bm{\theta}_{\text{Har}})}
    \end{aligned}
\end{equation}
where $\bm{\theta}_{\text{Geo}}$ are the phases that are maximizing the geometric mean and $\bm{\theta}_{\text{Har}}$ are the phases that are maximizing the harmonic mean. In addition to providing a performance bound between linear precoding and \ac{DPC}, this expression allows to evaluate the similarity of the \ac{GCEs}.
The bound ignores the fact, that orthogonal user channels (for which the performance gap is zero) can also arise for different channel eigenvalues (e.g. when $\bm{H}\bm{H}^\Her$ is diagonal with different diagonal entries).  Equal eigenvalues are therefore only a sufficient condition for channel orthogonality.\\

\section{RIS Optimization}\label{sec:RISOpt}
The maximization of the \ac{SE} at high \ac{SNR} for \ac{DPC} reads as [cf. \eqref{eq:HSNROffsets}]
\begin{equation}
    \underset{\bthet, \; \abs{\theta_n}=1\; \forall n}{\max} \log_2 \det(\bH \bH^\Her).
\end{equation}
The Gram channel matrix $\bH \bH^\Her$ can be rewritten as
\begin{equation}
    \begin{aligned}
        \bm{H}\bm{H}^{\He}   &=  (\bH_\td + \bH_\tre \mThet \bH_\ts)(\bH_\td + \bH_\tre \mThet \bH_\ts)^\Her\\
                        &= \bH_\td \bH_\td^\Her + \bH_\td  \bH_\ts^\Her \mThet^\Her \bH_\tre^\Her +  \bH_\tre \mThet \bH_\ts \bH_\td^\Her \\
                        &\quad+\bH_\tre \mThet \bH_\ts \bH_\ts^\Her \mThet^\Her \bH_\tre^\Her\\
                        &=  \bH_\td \bH_\td^\Her + \summe{l=1}{{R_\ts}} \big(\bH_\td   \sigma_l \bv_l \bu_l^\Her \mThet^\Her \bH_\tre^\Her \\
                        & \quad+  \bH_\tre \mThet \sigma_l \bu_l \bv_l^\Her\bH_\td^\Her+\bH_\tre \mThet \sigma_l^2 \bu_l\bu_l^\Her \mThet^\Her \bH_\tre^\Her\big)\\
                        &=  \bH_\td \bH_\td^\Her + \summe{l=1}{{R_\ts}} \left(\bD_l \bbthet \bbthet^\Her \bD_l^\Her - \bH_\td \bv_l (\bH_\td \bv_l)^\Her\right)\\
                        &=  \bH_\td\left(\eye - \summe{l=1}{{R_\ts}}\bv_l \bv_l^\Her\right)\bH_\td^\Her + \summe{l=1}{{R_\ts}} \bD_l \bbthet \bbthet^\Her \bD_l^\Her\\
                        &= \bC + \bQ
    \end{aligned}
\end{equation}
where $\bC =   \bH_\td\bm{P}^{R_\ts}\bH_\td^\Her$ and $\bQ =  \summe{l=1}{{R_\ts}} \bD_l \bbthet \bbthet^\Her \bD_l^\Her$.
The matrices $\bC$ and $\bQ$ are constructed with $\bD_l = \begin{bmatrix} \bH_\tre \diag(\bu_l)\sigma_l, \bH_\td \bv_l 
\end{bmatrix}$
and
$\bbthet^\Her = \begin{bmatrix}
    \bthet^\Her, 1
\end{bmatrix}$ by using the singular value decomposition $\bH_\ts = \summe{l=1}{{R_\ts}} \sigma_l \bu_l \bv_l^\Her$ and the orthogonal projector
$\bm{P}^{R_\ts} = \eye - \summe{l=1}{{R_\ts}}\bv_l \bv_l^\Her$.
The optimization problem now reads as
\begin{equation}\label{eq:GeoMeanMax}
    \begin{aligned}
        \underset{\bthet, \; \abs{\theta_n}={1} \; \forall n}{\max}\log_2\det(\bC + \bQ) \quad
        \text{s.t.} \quad \bQ = \summe{l=1}{{R_\ts}} \bD_l \bbthet \bbthet^\Her \bD_l^\Her.\\
    \end{aligned}
\end{equation}
If we could choose any positive semidefinite matrix $\bQ$ with $\rank(\bQ) \le R_\ts$ and a limited trace, the solution would result in rank-constrained waterfilling.
This was analyzed in \cite{RankCWF} and is equivalent to performing waterfilling over the ${R_\ts}$ smallest eigenvalues of the matrix $\bC$.\\
Even though we are additionally limited by the characteristics of the \ac{RIS} leading further restrictions on $\bQ$, we will see in the simulations that the solution shares similarities with the one of rank-constrained waterfilling. 
The resulting maximization under the unimodular constraints can be solved with rank relaxations similar to \ac{SDR} or with local optimal algorithms. In this article, the latter approach is chosen and we opt for high \ac{SNR} adjusted versions of the element-wise algorithms given in \cite{MIMOP2PCap},  \cite{MaxSumRateJour}.\\
Similarly, the maximization of the harmonic mean results in $  \arg \max \; r/\tr\left((\bH \bH^\Her)^{\inv} \right) = \arg \min \;\tr\left((\bC + \bQ)^{\inv} \right)$ under the same constraint as in \eqref{eq:GeoMeanMax}.
If we could choose again any positive semidefinite matrix $\bQ$ with $\rank(\bQ) \le R_\ts$ and a limited trace, the solution would also result in rank-constrained waterfilling which can be shown by combining the results of \cite{RankCWF} and \cite[Appendix B]{ChannelEst}. Even though similarities can be observed, under the additional constraints of the \ac{RIS}, the harmonic and geometric mean maximization will in general lead to different solutions.

 \section{Eigenvalue Bounds}
 We will now analyze the limitations of the eigenvalue modification w.r.t. the channel conditions. To this end, we start with a rank-one assumption of the channel $\bH_\ts$ and afterwards increase the rank to an arbitrary number.
\subsection{Rank-One Solution}
Assume that the channel between the \ac{BS} and the \ac{RIS} is \ac{LOS} dominated, e.g., when a \ac{mmWave} scenario is discussed. Even though this assumption is often used, we will see that it is quite limiting in view of the resulting degrees of freedom at the \ac{RIS}.
For the special case ${R_\ts}=1$, the Gram channel matrix reads as
\begin{equation}
    \begin{aligned}
    \bm{H}\bm{H}^{\He} &= \bm{C} + \bm{D}_1\bbthet \bbthet^\Her \bm{D}_1^\Her.\\
    \end{aligned}
\end{equation}
Introducing $\bW \bPhi \bW^\Her=\bm{C}$ as the \ac{EVD} (for all EVDs we assume a decreasing order of the eigenvalues) of $\bm{C}$, we know from the interlacing of eigenvalues (see for example {\cite[p.442, Theorem 8.5.3]{MatrixComp}}), that the eigenvalues of $\bH\bH^\Her$ interlace the ones of $\bC$  for any possible choice of $\bbthet$.\\
It follows, that the \ac{GCEs} of the direct channel, that is, when $\bthet = \bm{0}$, and the \ac{GCEs} for any choice of $\bthet$ can be bounded by 
\begin{equation}\label{eq:EWR1Interlacing}
    \phi_r \le \xi_r\le \phi_{r-1}\le \xi_{r-1}\le\dots  \le\phi_{1} \le  \xi_{1}
 \end{equation}
where $\xi_n$ are either the \ac{GCEs} $ \lambda_n^\td$ of $\bm{H}_\td \bm{H}_\td^\Her$ or the \ac{GCEs} $ \lambda_n$ of $\bm{H} \bm{H}^\Her$. We only have equality $ \phi_n = \xi_n$ if $\bm{e}_n^{\T} \bm{W}^\Her\bm{D}_1\bbthet = 0$ which means that the space of the rank-one update is orthogonal to $\mathrm{range}(\bm{w}_n)$ where $\bm{w}_n = \bm{W}\bm{e}_n$ is the eigenvector corresponding to the $n$-th eigenvalue.\\
It follows immediately from \eqref{eq:EWR1Interlacing} that the eigenvalue placement performed by the \ac{RIS} is restricted by
\begin{equation}\label{eq:RankOneLimit}
    \begin{aligned}
        \lambda_n & \le \lambda^\td_{n-1},  \; n=2,\dots,r\;
            & \text{and} \; \lambda_1 \; \text{being unbounded}.
    \end{aligned}
\end{equation}
The eigenvalues are of course further restricted by the maximum channel gain, i.e., $ \lambda_n \le \phi_n + \norm{\bm{D}_1 \bthet}_2^2$.
We can infer from \eqref{eq:RankOneLimit} that apart from the largest GCE, we can improve each GCE only up to the next larger one of the direct channel. \\
To see this more clearly, we consider a scenario in which a group of $G$ \ac{GCEs} are very small, i.e., $\lambda^\td_r \approx \lambda^\td_{r-1} \approx \dots \approx \lambda^\td_{r-G+1} \ll \lambda^\td_{r-G} \approx  \lambda^\td_{r-G-1} \approx \dots \approx \lambda^\td_1$ holds for the \ac{GCEs} of the direct channel.
In this case, the \ac{RIS} could only improve one (i.e., $\lambda^\td_{r-G+1}$) of the $G$ smaller \ac{GCEs} (to a maximum of approximately $\lambda^\td_{r-G}$). 
Moreover, if the \ac{RIS} has strong impact such that
$\lambda_r \approx \lambda^\td_{r-1},  \lambda_{r-1} \approx \lambda^\td_{r-2}, \dots,\lambda_2 \approx \lambda^\td_1$ 
holds, it can only improve the largest GCE $\lambda_1$ and the channel condition will get automatically worse. Note that when the influence of the \ac{RIS} is very strong, the transmission would just take place via the rank-one \ac{RIS} channel.
The same could be observed, if we had a perfectly conditioned direct channel (all \ac{GCEs} are approximately equal).
Then the \ac{RIS} could only worsen the channel condition by improving the largest GCE.\\
It is important to note that even if the channel $\bm{H}_\ts$ has only a rank of one, the \ac{RIS} does not deteriorate the performance.
The \ac{SE} will increase monotonically with a rank-one \ac{BS}-\ac{RIS} channel.
However, the performance will be signifcantly degraded in comparison to a multi-rank \ac{BS}-\ac{RIS} channel as only a single eigenvalue can be increased without limitations.
\subsection{Multi-Rank Solution}\label{sec:MultiRank}
The eigenvalue limitations soften when the $\rank(\bm{H}_\ts) = R_\ts$ increases. Similar to the rank-one discussion, having a total of ${R_\ts}$ rank-one updates available, one could improve the eigenvalues up to the next ${R_\ts}$ larger eigenvalues.
Considering the rank-improvement example from above, ${R_\ts}$ of the $G$ eigenvalues could be improved in this scenario. 
Additionally, if the impact of the \ac{RIS} is large, the ${R_\ts}$ largest eigenvalues can be further improved. \\
We can therefore conclude that if ${R_\ts}\ge r$, all eigenvalues can be controlled and a perfectly conditioned channel (also in the extreme case, when the transmission would effectively only take place via the \ac{RIS}) can be obtained.
\subsection{Multi-RIS Scenario}
Typically, strong \ac{LOS} conditions for the \ac{BS}-\ac{RIS} channel are beneficial as can be seen by e.g. the wide-band considerations in \cite{Wideband}.
To additionally guarantee a high rank of $\bm{H}_\ts$, the deployment of multiple RISs is necessary.
Consequently, the maximum number of controllable eigenvalues is increased to
\begin{equation}
    \#\text{EVs} = \summe{n=1}{N_{\text{RISs}}}  \rank(\bm{H}_{\ts,n}).
\end{equation}
The multi-\ac{RIS} scenario as a possible solution to the above eigenvalue limitations will be analyzed in a future publication.

\section{Simulations}
We consider a scenario similar to the one of \cite{WSR} and \cite{IRSLISA}, where $6$ single antenna ($\NM=1$) users are uniformly distributed in a circle  with radius $10\,\text{m}$ centered at ($200\,\text{m}$, $30\,\text{m}$) and served by an $\NB=16$ antenna base station located at ($0\,\text{m}$, $0\,\text{m}$).
The transmission is enhanced by an \ac{RIS} located at ($200\,\text{m}$, $0\,\text{m}$). The path losses for the channels are assumed to follow the model $ L_{\text{dB}} = \alpha + \beta 10\log_{10}(\frac{d}{\mathrm{m}})$ (we assume $\alpha_\td = \alpha_\tre = \alpha_\ts=30\,\mathrm{dB}$ for $\bH_\tdk, \bH_\trek ,\; \text{and}\; \bH_\ts$ in all simulations)
where $d$ is the distance between the receiver and the transmitter. For all plots, $1000$ realizations 
and a noise variance of $\sigma^2=-100\,\text{dBm}$ is assumed at each receive antenna.
Furthermore, $\beta_\td=3.76$ and uncorrelated Rayleigh fading is assumed for $\bH_\tdk$ if not stated otherwise.
The parameters $\beta_\tre$ and $\beta_\ts$ will be defined below.
\subsection{Rank Constrained Waterfilling}
At first, we will show numerically how the eigenvalues are affected by the \ac{RIS} optimization.
To see the impact on the eigenvalues more clearly, we introduce an extra path loss of $20\,\text{dB}$ for 3 of the 6 users.
Furthermore, the channel $\bH_\trek$ is modeled as uncorrelated Rayleigh fading with $\beta_\tre=3.76$.
To see the importance of $\rank(\bH_\ts)$, we model $\bH_\ts$ with the Kronecker channel model as
    $\bH_\ts = \sqrt{L_\ts} \sqrt{\frac{\NB}{R_\ts}}\bm{M} \left[\begin{smallmatrix}
        \;\;\eye_{R_\ts}&\bm{0}\\
        \bm{0}& \bm{0}_{\NB-R_\ts}
    \end{smallmatrix}\right]\bm{S}^\Her$
to analyze the behavior for a different rank. The elements of the matrix $\bm{M}$ are distributed as $\gaussdist{0}{1}$ and the unitary matrix $\bm{S}$ is chosen by computing the QR decomposition of a random matrix where all elements are distributed as $\gaussdist{0}{1}$.
The channel gain normalization $\sqrt{\frac{\NB}{R_\ts}}$ is introduced as we would like to analyze the system only w.r.t. the structure (rank) of the matrix $\bH_s$.
The pathloss parameter for this channel is chosen as  $\beta_\ts = 2.2$.
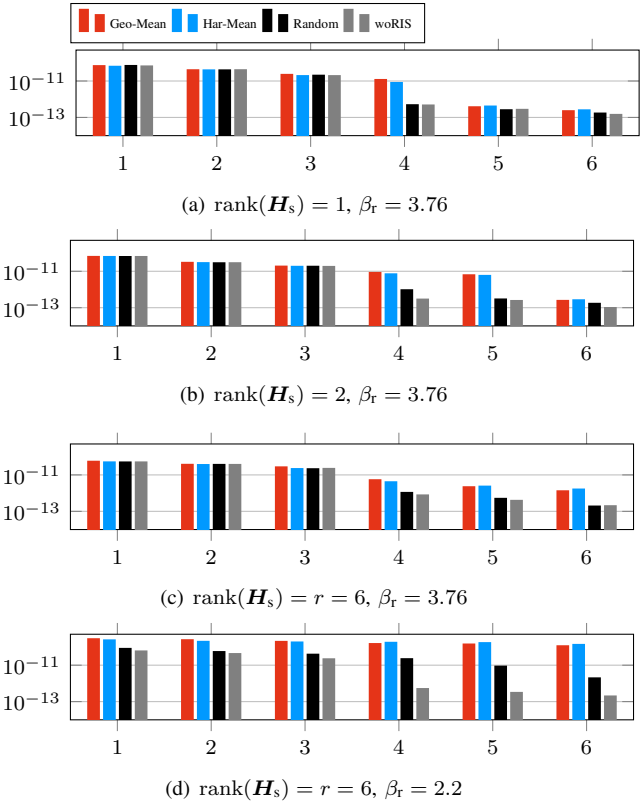
\begin{figure}[h!]
    \centering
    \subfigure[$\rank(\bH_\ts) = 1$, $\beta_\tre=3.76$]{

        \pgfplotstableread{plot_data/Eigs1.txt}{\tableEigs}

        \begin{tikzpicture}
            \begin{axis}[
                ticklabel style = {font=\footnotesize},
                xtick = data,
                width = 0.5\textwidth,
                height = 0.15\textwidth,
                ymin=1e-14,
                ymax =5e-10,
                ymode=log,
                ymajorgrids,
                log origin=infty,
                ybar,
                bar width = 4pt,
                legend style={at={(0.62,1.55)}},
                legend style={font=\tiny},
                legend columns=4,
                anchor=south east,
            ]
                \addplot[TUMBeamerRed,thick,fill=TUMBeamerRed] table [x={N}, y = {HSNRDPC}] {\tableEigs};
                \addplot[TUMBeamerBlue,thick,fill=TUMBeamerBlue] table [x={N}, y = {HSNRLinear}] {\tableEigs};
                \addplot[black,thick,fill=black] table [x={N}, y = {HSNRLinearRandom}] {\tableEigs};
                \addplot[gray,thick,fill=gray] table [x={N}, y = {HSNRLinearWO}] {\tableEigs};

            \legend{Geo-Mean,Har-Mean,Random,woRIS}
            \end{axis}
            \end{tikzpicture}
            \label{fig:Eigsa}

    }
    \subfigure[$\rank(\bH_\ts) = 2$, $\beta_\tre=3.76$]{

        \pgfplotstableread{plot_data/Eigs2.txt}{\tableEigs}

        \begin{tikzpicture}
            \begin{axis}[
                ticklabel style = {font=\footnotesize},
                xtick = data,
                width = 0.5\textwidth,
                height = 0.15\textwidth,
                ymin=1e-14,
                ymax =5e-10,
                ymode=log,
                ymajorgrids,
                log origin=infty,
                ybar,
                bar width = 4pt,
                legend style={at={(1,1)}},
                legend columns=4
            ]
                \addplot[TUMBeamerRed,thick,fill=TUMBeamerRed] table [x={N}, y = {HSNRDPC}] {\tableEigs};
                \addplot[TUMBeamerBlue,thick,fill=TUMBeamerBlue] table [x={N}, y = {HSNRLinear}] {\tableEigs};
                \addplot[black,thick,fill=black] table [x={N}, y = {HSNRLinearRandom}] {\tableEigs};
                \addplot[gray,thick,fill=gray] table [x={N}, y = {HSNRLinearWO}] {\tableEigs};
    
            \end{axis}
            \end{tikzpicture}
            \label{fig:Eigsb}

    }

    \subfigure[$\rank(\bH_\ts) = r = 6$, $\beta_\tre=3.76$]{

        \pgfplotstableread{plot_data/Eigs6.txt}{\tableEigs}

        \begin{tikzpicture}
            \begin{axis}[
                ticklabel style = {font=\footnotesize},
                xtick = data,
                width = 0.5\textwidth,
                height = 0.15\textwidth,
                ymin=1e-14,
                ymax =5e-10,
                ymode=log,
                ymajorgrids,
                log origin=infty,
                ybar,
                bar width = 4pt,
                legend style={at={(1,1)}},
                legend columns=4
            ]
                \addplot[TUMBeamerRed,thick,fill=TUMBeamerRed] table [x={N}, y = {HSNRDPC}] {\tableEigs};
                \addplot[TUMBeamerBlue,thick,fill=TUMBeamerBlue] table [x={N}, y = {HSNRLinear}] {\tableEigs};
                \addplot[black,thick,fill=black] table [x={N}, y = {HSNRLinearRandom}] {\tableEigs};
                \addplot[gray,thick,fill=gray] table [x={N}, y = {HSNRLinearWO}] {\tableEigs};    
            \end{axis}
            \end{tikzpicture}
            \label{fig:Eigsc}

    }
    \subfigure[$\rank(\bH_\ts) = r = 6$, $\beta_\tre=2.2$]{

        \pgfplotstableread{plot_data/Eigs6H.txt}{\tableEigs}

        \begin{tikzpicture}
            \begin{axis}[
                ticklabel style = {font=\footnotesize},
                xtick = data,
                width = 0.5\textwidth,
                height = 0.15\textwidth,
                ymin=1e-14,
                ymax =5e-10,
                ymode=log,
                ymajorgrids,
                log origin=infty,
                ybar,
                bar width = 4pt,
                legend style={at={(1,1)}},
                legend columns=4
            ]
                \addplot[TUMBeamerRed,thick,fill=TUMBeamerRed] table [x={N}, y = {HSNRDPC}] {\tableEigs};
                \addplot[TUMBeamerBlue,thick,fill=TUMBeamerBlue] table [x={N}, y = {HSNRLinear}] {\tableEigs};
                \addplot[black,thick,fill=black] table [x={N}, y = {HSNRLinearRandom}] {\tableEigs};
                \addplot[gray,thick,fill=gray] table [x={N}, y = {HSNRLinearWO}] {\tableEigs};
    
            \end{axis}
            \end{tikzpicture}
            \label{fig:Eigsd}

    }
    
    \caption{\ac{GCEs} for different ranks of $\bH_\ts$ and different gains of the reflective channel $\bH_\trek$.}
    \label{fig:Eigs}
    \vspace{-10pt}
 \end{figure} 
 In Figure \ref{fig:Eigs}, the eigenvalues after the \ac{RIS} optimization can be seen.
 For the geometric mean optimization (Geo-Mean) a modified version of \cite{MaxSumRateJour} was used.
 The harmonic mean maximization was performed with the algorithm presented in \cite{IRSLISA}.
Both algorithms were initialized with the same random phase shifts (uniformly distributed phases) which are also presented in Figure \ref{fig:Eigs} (Random).\\
With $\bH_\ts$ having a rank of one [see Figure \ref{fig:Eigsa}], it can be seen that the discussed limitation in \eqref{eq:RankOneLimit} $  \lambda_n \le \lambda_{n-1}^\td \quad \forall n=2,\dots,6$ holds.
Therefore, only one of the 3 smaller eigenvalues could be improved. \\
Increasing the rank of $\bH_\ts$ to two and six results in Figure \ref{fig:Eigsb} and Figure \ref{fig:Eigsc}, respectively. At first, two and then all three of the smaller eigenvalues could be improved which is in accordance with the discussion in subsection \ref{sec:MultiRank}.
Additionally, the channel gain is spread over the controlled eigenvalues which highlights the similarities to the rank-constrained waterfilling solution mentioned in Section \ref{sec:RISOpt}.\\
Furthermore, we can see in Figure \ref{fig:Eigsd} that if the \ac{RIS} has enough impact ($\beta_\tre=2.2$), then a rank of $\rank(\bH_\ts)=r=6$ is enough to control all eigenvalues and to obtain a well-conditioned channel.
 \subsection{Performance Analysis with multi-antenna Users}
We will now analyze the behavior of the eigenvalues connected with the performance of \ac{DPC} and linear precoding.
In comparison to the last subsection, we consider $K=3$ users with $\NM=2$ antennas each instead of the 6 single-antenna users.
The channel $\bH_\trek$ is assumed to follow uncorrelated Rayleigh fading with a channel gain of $\beta_\tre=2.2$.
Instead of assuming  $\bH_\ts$ is rank one, we assume a more practical assumption in which  $\bH_\ts$ is assumed to follow Rician fading.
Particularly, a Rician component with a Rician factor of $10\,\text{dB}$ is considered in the following.
The Rician component is obtained by assuming a half-wavelength \ac{ULA} at the \ac{BS} and at the \ac{RIS} where the \ac{AoA} and \ac{AoD} are selected uniformly from the interval $[0,2\pi)$.
This model is then compared to $\bH_\ts$ being Rayleigh (uncorrelated) distributed. 
 \begin{figure}[h!]
    \centering
    \subfigure[$\bH_\ts$ Rayleigh fading]{
      \pgfplotstableread{plot_data/EigsHGFR.txt}{\tableEigs}

        \begin{tikzpicture}
            \begin{axis}[
                ticklabel style = {font=\footnotesize},
                xtick = data,
                width = 0.5\textwidth,
                height = 0.15\textwidth,
                ymin = 5e-13,
                ymax = 5e-9,
                ymode=log,
                ymajorgrids,
                log origin=infty,
                ybar,
                bar width = 4pt,
                legend style={at={(0.62,1.55)}},
                legend style={font=\tiny},
                legend columns=4,
                anchor=north west,
            ]
                \addplot[TUMBeamerRed,thick,fill=TUMBeamerRed] table [x={N}, y = {HSNRDPC}] {\tableEigs};
                \addplot[TUMBeamerBlue,thick,fill=TUMBeamerBlue] table [x={N}, y = {HSNRLinear}] {\tableEigs};
                \addplot[black,thick,fill=black] table [x={N}, y = {HSNRLinearRandom}] {\tableEigs};
                \addplot[gray,thick,fill=gray] table [x={N}, y = {HSNRLinearWO}] {\tableEigs};

                \legend{Geo-Mean,Har-Mean,Random,woRIS}
            \end{axis}
            \end{tikzpicture}
            \label{fig:PerfEigsa}
    }
    \subfigure[$\bH_\ts$ $10\,\text{dB}$ Rician fading]{
        \pgfplotstableread{plot_data/EigsRank1.txt}{\tableEigs}
            \begin{tikzpicture}
                \begin{axis}[
                    ticklabel style = {font=\footnotesize},
                    xtick = data,
                    ymin = 5e-13,
                    ymax = 5e-9,
                    width = 0.5\textwidth,
                    height = 0.15\textwidth,
                    ymode=log,
                    ymajorgrids,
                    log origin=infty,
                    ybar,
                    bar width = 4pt,
                    legend style={at={(1,1)}},
                    legend columns=4
                ]
                  \addplot[TUMBeamerRed,thick,fill=TUMBeamerRed] table [x={N}, y = {HSNRDPC}] {\tableEigs};
                  \addplot[TUMBeamerBlue,thick,fill=TUMBeamerBlue] table [x={N}, y = {HSNRLinear}] {\tableEigs};
                  \addplot[black,thick,fill=black] table [x={N}, y = {HSNRLinearRandom}] {\tableEigs};
                  \addplot[gray,thick,fill=gray] table [x={N}, y = {HSNRLinearWO}] {\tableEigs};

                \end{axis}
                \end{tikzpicture}
                \label{fig:PerfEigsb}
    }
    \caption{\ac{GCEs} for the Rayleigh and Rician fading model of the matrix $\bH_\ts$ with a power of $P=40\,\text{dBm}$ and $\NRe = 256$ reflecting elements.}
    \label{fig:PerfEigs}
\vspace{-5pt}
            \end{figure}
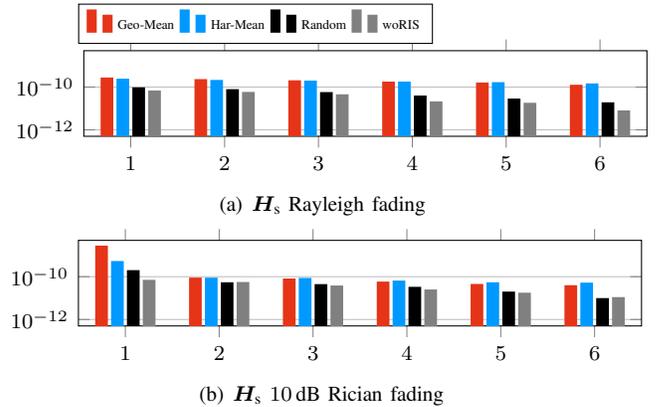\newline
At first, we analyze the behavior of the eigenvalues shown in Figure \ref{fig:PerfEigs}.
For Rayleigh fading [see Figure \ref{fig:PerfEigsa}] all eigenvalues can be controlled resulting in a well-conditioned channel.
In the case of Rician fading [see Figure \ref{fig:PerfEigsb}], the channel has full rank as well and the smaller eigenvalues can also be improved.
However, we can already see the impact of the rank-one component resulting in the increase of the largest eigenvalue.
\pgfplotstableread{plot_data/SNRRates.txt}{\tableSNR}
\begin{figure}[h!]
    \flushleft
    \hspace*{-5pt}
    \subfigure[$\bH_\ts$ Rayleigh fading]{
\begin{tikzpicture}
        \begin{axis}[
         ticklabel style = {font=\footnotesize},
        xmin = -10, xmax = 40,
        ymin = 0, ymax =70,
        xtick = {0,20,40},
        grid = both,
        minor tick num = 1,
        major grid style = {lightgray},
        minor grid style = {lightgray!25},
        width = 0.28\textwidth,
        height = 0.25\textwidth,
        xlabel = {\footnotesize{Power [dBm]}},
        ylabel = {\footnotesize{\ac{SE}  [bpcu]}},
        x label style={at={(\textwidth* 0.28*0.35,0.07)},anchor=north, below=0mm},
        y label style={at={(0.12,\textwidth* 0.25*0.3)},anchor=north, below=0mm},
        legend pos = north west,
        legend cell align=left,
        reverse legend,
        legend columns = 1,
        legend style={font=\tiny},
        legend style={at={(-0.03,1.0)}},
        ]
     
        \addplot[gray, dotted,line width=1pt ] table [x = {Power}, y = {HSNRLinearWO}] {\tableSNR};
        \addplot[black,dashed] table [x = {Power}, y = {HSNRDPCWO}] {\tableSNR};

        \addplot[TUMBeamerGreen, dotted,line width=1pt] table [x = {Power}, y = {HSNRLinearRandom}] {\tableSNR};
        \addplot[TUMBeamerOrange, dashed] table [x = {Power}, y = {HSNRDPCRandom}] {\tableSNR};
        
        \addplot[TUMBeamerBlue,dotted,line width=1pt] table [x = {Power}, y = {HSNRLinear}] {\tableSNR};
        \addplot[TUMBeamerRed,dashed] table [x = {Power}, y = {HSNRDPC}] {\tableSNR};

        \addplot[gray, mark = diamond] table [x = {Power}, y = {LISAwoIRS}] {\tableSNR};
        \addplot[black, mark = star] table [x = {Power}, y = {DPCwoIRS}] {\tableSNR};

        \addplot[TUMBeamerGreen, mark = triangle] table [x = {Power}, y = {LISARandom}] {\tableSNR};
        \addplot[TUMBeamerOrange, mark = otimes] table [x = {Power}, y = {DPCRandom}] {\tableSNR};
        
        \addplot[TUMBeamerBlue, mark = *] table [x = {Power}, y = {IRSLISA}] {\tableSNR};
        \addplot[TUMBeamerRed, mark = square] table [x = {Power}, y = {DPC-AO}] {\tableSNR};

        \legend{LISAwoRIS,
                DPCwoRIS,
                LISARandom,
                DPCRandom,
                RIS-LISA,
                DPC-AO}

    \end{axis}

    \end{tikzpicture}
    \label{fig:SNRCompa}
    }\hspace*{-10pt}\subfigure[$\bH_\ts$ $10\,\text{dB}$ Rician fading]
    {
        \pgfplotstableread{plot_data/SNRRatesRank1.txt}{\tableSNR}
        \begin{tikzpicture}
                \begin{axis}[
                    ticklabel style = {font=\footnotesize},
                 xmin = -10, xmax = 40,
                ymin = 0, ymax = 70,
                xtick = {0,20,40},
                grid = both,
                minor tick num = 1,
                major grid style = {lightgray},
                minor grid style = {lightgray!25},
                width = 0.28\textwidth,
                height = 0.25*\textwidth,
                xlabel = {\footnotesize{Power [dBm]}},
                x label style={at={(\textwidth* 0.28*0.35,0.07)},anchor=north, below=0mm},
                y label style={at={(0.12,\textwidth* 0.25*0.3)},anchor=north, below=0mm},
                ylabel = {\footnotesize{\ac{SE} [bpcu]}},
                legend pos = north west,
                reverse legend,
                legend style={at={(-0.1,1.25)}},
                legend style={font=\tiny},
                legend columns=3,
                anchor=north west]
             
                \addplot[gray, dotted,line width=1pt ] table [x = {Power}, y = {HSNRLinearWO}] {\tableSNR};
                \addplot[black,dashed] table [x = {Power}, y = {HSNRDPCWO}] {\tableSNR};
        
                \addplot[TUMBeamerGreen, dotted,line width=1pt] table [x = {Power}, y = {HSNRLinearRandom}] {\tableSNR};
                \addplot[TUMBeamerOrange, dashed] table [x = {Power}, y = {HSNRDPCRandom}] {\tableSNR};
                
                \addplot[TUMBeamerBlue,dotted,line width=1pt] table [x = {Power}, y = {HSNRLinear}] {\tableSNR};
                \addplot[TUMBeamerRed,dashed] table [x = {Power}, y = {HSNRDPC}] {\tableSNR};

                \addplot[gray, mark = diamond] table [x = {Power}, y = {LISAwoIRS}] {\tableSNR};
                \addplot[black, mark = star] table [x = {Power}, y = {DPCwoIRS}] {\tableSNR};
        
                \addplot[TUMBeamerGreen, mark = triangle] table [x = {Power}, y = {LISARandom}] {\tableSNR};
                \addplot[TUMBeamerOrange, mark = otimes] table [x = {Power}, y = {DPCRandom}] {\tableSNR};
                
                \addplot[TUMBeamerBlue, mark = *] table [x = {Power}, y = {IRSLISA}] {\tableSNR};
                \addplot[TUMBeamerRed, mark = square] table [x = {Power}, y = {DPC-AO}] {\tableSNR};

            \end{axis}

            \end{tikzpicture}
            \label{fig:SNRCompb}
    }
    \caption{Comparison of the \ac{SE} with $\NRe=256$ \ac{RIS} Elements for the Rayleigh and Rician fading channel model of $\bH_\ts$.}
    \label{fig:SNRComp}
    \vspace{-5pt}
\end{figure}
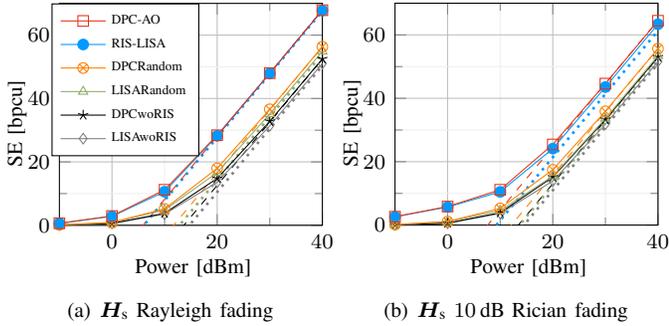 
This channel characteristic directly affects the obtainable \ac{SE} as seen in Figure \ref{fig:SNRComp} where the performance of \ac{DPC} and linear precoding is compared.
For DPC-AO, we use the algorithm from \cite{MaxSumRateJour} whereas for linear precoding schemes, we use the RIS-LISA algorithm proposed in \cite{IRSLISA}.\\
When considering only the direct channel and the random phase shifts, a perfomance gap can be observed between the linear schemes and \ac{DPC}. In case of DPC-AO and RIS-LISA (optimized phase shifts), the performance gap only remains in the case of Rician Fading [see Figure \ref{fig:SNRCompb}] whereas for Rayleigh fading [see Figure \ref{fig:SNRCompa}] the methods perform similarly.
This can be explained by analyzing the harmonic mean (dotted lines) and the geometric mean (dashed lines) based high \ac{SNR} approximations. Please note, that for linear precoding, the lower bound in \eqref{eq:HarmonicMeanBound} instead of the exact approximation in \eqref{eq:HSNROffsets} is used which allows an interpretation based on the \ac{GCEs}. This means that in the high \ac{SNR} region, \ac{RIS}-LISA will always lie in between the mentioned lower bound and the high \ac{SNR} approximation of \ac{DPC}.
For Rayleigh fading, the eigenvalues are similar for the optimized phase shifts and the gap between the high \ac{SNR} approximations vanishes. It directly follows that \ac{RIS}-LISA and \ac{DPC}-AO result in a comparable performance at high \ac{SNR}. 
\pgfplotstableread{plot_data/IRSElRates.txt}{\tableIRSEl}
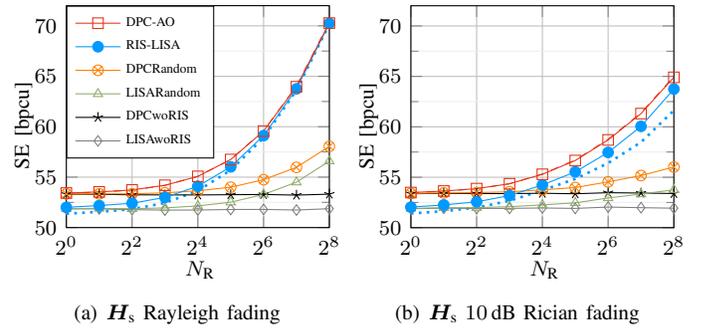
\begin{figure}[h!]
    \centering
    \hspace*{-5pt}\subfigure[$\bH_\ts$ Rayleigh fading]{
\begin{tikzpicture}
    \begin{axis}[
        ticklabel style = {font=\footnotesize},
    xmin = 1, xmax = 256,
    ymin = 50, ymax = 72,
    xtick = {1,4,16,64,256},
    xmode = log,
    log basis x = {2},
    grid = both,
    minor tick num = 1,
    major grid style = {lightgray},
    minor grid style = {lightgray!25},
    width = 0.28\textwidth,
    height = 0.25*\textwidth,
    xlabel = {\footnotesize{$\NRe$}},
    x label style={at={(\textwidth* 0.28*0.35,0.07)},anchor=north, below=0mm},
    y label style={at={(0.12,\textwidth* 0.25*0.3)},anchor=north, below=0mm},
    ylabel = {\footnotesize{\ac{SE} [bpcu]}},
    legend pos = north west,
    reverse legend,
    legend cell align=left,
    legend columns = 1,
    legend style={font=\tiny},
    legend style={at={(0.,1.0)}}]

        \addplot[TUMBeamerBlue,dotted,line width=1pt] table [x = {N}, y = {HSNRLinear}] {\tableIRSEl};
        \addplot[TUMBeamerRed,dashed] table [x = {N}, y = {HSNRDPC}] {\tableIRSEl};

        \addplot[gray, mark = diamond] table [x = {N}, y = {LISAwoIRS}] {\tableIRSEl};
        \addplot[black, mark = star] table [x = {N}, y = {DPCwoIRS}] {\tableIRSEl};

        \addplot[TUMBeamerGreen, mark = triangle] table [x = {N}, y = {LISARandom}] {\tableIRSEl};
        \addplot[TUMBeamerOrange, mark = otimes] table [x = {N}, y = {DPCRandom}] {\tableIRSEl};
        
        \addplot[TUMBeamerBlue, mark = *] table [x = {N}, y = {IRSLISA}] {\tableIRSEl};
        \addplot[TUMBeamerRed, mark = square] table [x = {N}, y = {DPC-AO}] {\tableIRSEl};

        \legend{LISAwoRIS,
                DPCwoRIS,
                LISARandom,
                DPCRandom,
                RIS-LISA,
                DPC-AO}

    \end{axis}
     
    \end{tikzpicture}
    \label{fig:RefCompa}
    }\hspace*{-10pt}\subfigure[$\bH_\ts$ $10\,\text{dB}$ Rician fading]{

        \pgfplotstableread{plot_data/IRSElRatesRank1.txt}{\tableIRSEl}
        \begin{tikzpicture}
            \begin{axis}[
                ticklabel style = {font=\footnotesize},
            xmin = 1, xmax = 256,
            ymin = 50, ymax = 72,
            xtick = {1,4,16,64,256},
            xmode = log,
            log basis x = {2},
            grid = both,
            minor tick num = 1,
            major grid style = {lightgray},
            minor grid style = {lightgray!25},
            width = 0.28\textwidth,
            height = 0.25*\textwidth,
            xlabel = {\footnotesize{$\NRe$}},
            x label style={at={(\textwidth* 0.28*0.35,0.07)},anchor=north, below=0mm},
            y label style={at={(0.12,\textwidth* 0.25*0.3)},anchor=north, below=0mm},
            ylabel = {\footnotesize{\ac{SE} [bpcu]}},
            legend pos = north west,
            reverse legend,
            legend style={at={(-0.03,1.0)}},
            legend style={font=\tiny},
            legend columns=3,
            anchor=north west]

                \addplot[TUMBeamerBlue,dotted,line width=1pt] table [x = {N}, y = {HSNRLinear}] {\tableIRSEl};
                \addplot[TUMBeamerRed,dashed] table [x = {N}, y = {HSNRDPC}] {\tableIRSEl};

                \addplot[gray, mark = diamond] table [x = {N}, y = {LISAwoIRS}] {\tableIRSEl};
                \addplot[black, mark = star] table [x = {N}, y = {DPCwoIRS}] {\tableIRSEl};
        
                \addplot[TUMBeamerGreen, mark = triangle] table [x = {N}, y = {LISARandom}] {\tableIRSEl};
                \addplot[TUMBeamerOrange, mark = otimes] table [x = {N}, y = {DPCRandom}] {\tableIRSEl};
                
                \addplot[TUMBeamerBlue, mark = *] table [x = {N}, y = {IRSLISA}] {\tableIRSEl};
                \addplot[TUMBeamerRed, mark = square] table [x = {N}, y = {DPC-AO}] {\tableIRSEl};
        
            \end{axis}
             
            \end{tikzpicture}       
            \label{fig:RefCompb}

    }
    \caption{Comparison of the \ac{SE} with a power of $P=40\,\mathrm{dBm}$ for the Rayleigh and Rician fading channel model of $\bH_\ts$.}
    \label{fig:RefComp}
\vspace{-5pt}
\end{figure}
Analyzing this behavior in greater depth results in Figure \ref{fig:RefComp}.
We can see that the gap between \ac{DPC} and linear precoding stays approximately constant for an increasing number of \ac{RIS} elements when random phase shifts are considered.
For the DPC-AO and RIS-LISA algorithms a different behavior can be observed. In the case of Rayleigh fading [see Figure \ref{fig:RefCompa}], the gap between the two schemes vanishes with an increasing number of elements whereas for Rician fading [see Figure \ref{fig:RefCompb}] the gap remains also for a higher number of elements. While the eigenvalue placement is limited (seen by the high \ac{SNR} approximations), the difference between the two methods can still decrease as favorable propagation will result in orthogonal user channels for large array sizes. Furthermore, both methods perform worse in case of Rician fading because of the eigenvalue limitations resulting in restrictions for spatial multiplexing.
\pgfplotstableread{plot_data/RicianRates.txt}{\tableIRSEl}
\begin{figure}[h!]
    \centering
    \hspace*{-5pt}\subfigure[$\beta_\td = 3.76$ ]{
\begin{tikzpicture}
    \begin{axis}[
        ticklabel style = {font=\footnotesize},
    xmin = -40, xmax = 80,
    ymin = 50, ymax = 70,
    xtick = {-40,0,40,80},
    grid = both,
    minor tick num = 1,
    major grid style = {lightgray},
    minor grid style = {lightgray!25},
    width = 0.28\textwidth,
    height = 0.25*\textwidth,
    xlabel = {\footnotesize{Rician factor of $\bm{H}_\ts$ [dB]}},
    x label style={at={(\textwidth* 0.28*0.35,0.07)},anchor=north, below=0mm},
    y label style={at={(0.12,\textwidth* 0.25*0.3)},anchor=north, below=0mm},
    ylabel = {\footnotesize{\ac{SE} [bpcu]}},
    legend pos = north west,
    reverse legend,
    legend columns = 1,
    legend style={font=\tiny},
    legend style={at={(0.,1.0)}}]

        \addplot[gray, mark = diamond] table [x = {Rician}, y = {LISAwoIRS}] {\tableIRSEl};
        \addplot[black, mark = star] table [x = {Rician}, y = {DPCwoIRS}] {\tableIRSEl};

        \addplot[TUMBeamerGreen, mark = triangle] table [x = {Rician}, y = {LISARandom}] {\tableIRSEl};
        \addplot[TUMBeamerOrange, mark = otimes] table [x = {Rician}, y = {DPCRandom}] {\tableIRSEl};
        
        \addplot[TUMBeamerBlue, mark = *] table [x = {Rician}, y = {IRSLISA}] {\tableIRSEl};
        \addplot[TUMBeamerRed, mark = square] table [x = {Rician}, y = {DPC-AO}] {\tableIRSEl};

        \addplot[TUMBeamerBlue,dotted,line width=1pt] table [x = {Rician}, y = {HSNRLinear}] {\tableIRSEl};
        \addplot[TUMBeamerRed,dashed] table [x = {Rician}, y = {HSNRDPC}] {\tableIRSEl};

    \end{axis}
     
    \end{tikzpicture}
    \label{fig:RicCompa}
    }\hspace*{-10pt}\subfigure[$\beta_\td = 5.76$  ]{

        \pgfplotstableread{plot_data/RicianRatesNoDirect.txt}{\tableIRSEl}
        \begin{tikzpicture}
                \begin{axis}[
                    ticklabel style = {font=\footnotesize},
                xmin = -40, xmax = 80,
                ymin = 0, ymax = 70,
                xtick = {-40,0,40,80},
                grid = both,
                minor tick num = 1,
                major grid style = {lightgray},
                minor grid style = {lightgray!25},
                width = 0.28\textwidth,
                height = 0.25*\textwidth,
                xlabel = {\footnotesize{Rician factor of $\bm{H}_\ts$ [dB]}},
                x label style={at={(\textwidth* 0.28*0.35,0.07)},anchor=north, below=0mm},
                y label style={at={(0.12,\textwidth* 0.25*0.3)},anchor=north, below=0mm},
                ylabel = {\footnotesize{\ac{SE} [bpcu]}},
                legend pos = north west,
                reverse legend,
                legend columns = 1,
                legend style={font=\tiny},
                legend style={at={(0.,1.0)}}]

                \addplot[TUMBeamerBlue,dotted,line width=1pt] table [x = {Rician}, y = {HSNRLinear}] {\tableIRSEl};
                \addplot[TUMBeamerRed,dashed] table [x = {Rician}, y = {HSNRDPC}] {\tableIRSEl};

                \addplot[gray, mark = diamond] table [x = {Rician}, y = {LISAwoIRS}] {\tableIRSEl};
                \addplot[black, mark = star] table [x = {Rician}, y = {DPCwoIRS}] {\tableIRSEl};
        
                \addplot[TUMBeamerGreen, mark = triangle] table [x = {Rician}, y = {LISARandom}] {\tableIRSEl};
                \addplot[TUMBeamerOrange, mark = otimes] table [x = {Rician}, y = {DPCRandom}] {\tableIRSEl};
                
                \addplot[TUMBeamerBlue, mark = *] table [x = {Rician}, y = {IRSLISA}] {\tableIRSEl};
                \addplot[TUMBeamerRed, mark = square] table [x = {Rician}, y = {DPC-AO}] {\tableIRSEl};

            \end{axis}
             
            \end{tikzpicture}       
            \label{fig:RicCompb}

    }
    \caption{Comparison of the \ac{SE} with a power of $P=40\mathrm{dBm}$ and $\NRe=256$ reflecting elements for a different impact of the direct channel.}
    \label{fig:RicComp}

\end{figure}
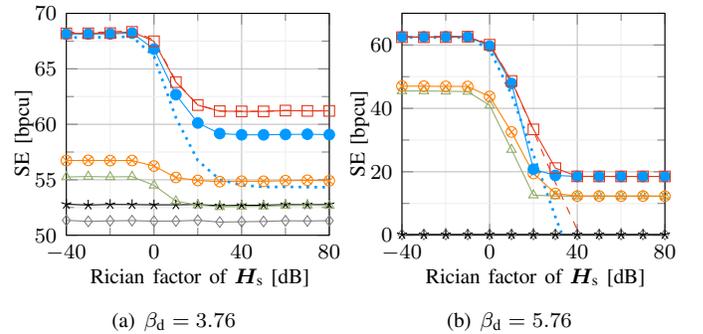
These limitations can be further observed in Figure \ref{fig:RicComp}. We can see that an increasing Rician factor deteriorates the performance for both random and optimized phases due to the eigenvalue limitations. Additionally, the high-\ac{SNR} approximations are not valid anymore if the Rician factor is large. This is especially pronounced for a negligible direct channel [see Figure \ref{fig:RicCompb}]. The circumstance that one eigenvalue is dominant leads to both \ac{DPC} and \ac{RIS}-LISA allocating only a single stream resulting in the same performance. The transmit power would have to be increased dramatically to match the high-\ac{SNR} approximations.

\section{Conclusion}
We have seen that the \ac{RIS} optimization directly affects the placement of the channel eigenvalues by introducing a connection with the geometric and harmonic mean.
In particular, we have seen that if the rank of the channel between the BS and the RIS is high, we can control all eigenvalues resulting in a good channel condition and the gap between linear precoding schemes and DPC vanishes.
On the contrary, when the BS-RIS channel has low rank (or even only LOS), the capabilities of the \ac{RIS} are clearly limited. This can be circumvented by considering multiple RISs.

\bibliographystyle{IEEEtran}
\bibliography{refs}
\end{document}